\documentstyle[epsf]{article}
\newcommand{\lsim}{\mbox{\raisebox{-.3em}{$\stackrel{<}{\sim}$}}}
\renewcommand{\cite}[1]{\ref{#1}}
\newcommand{\beq}{\begin{equation}}
\newcommand{\eeq}{\end{equation}}
\newcommand{\beqa}{\begin{eqnarray}}
\newcommand{\eeqa}{\end{eqnarray}}
\newcommand{\lmd}{\lambda}

\begin{document}
\baselineskip=0.6cm
\mbox{}\\[-3.5em]

\begin{center}
{\Large\bf Re/Os constraint on the time-variability of \\
the fine-structure constant}\\[.6em]
Yasunori Fujii$^1$ and Akira Iwamoto$^2$\\[.6em]
{\small
\hspace*{-.4em}$^1$Advanced \hspace{-.2em}Research \hspace{-.2em}Institute \hspace{-.2em}for \hspace{-.2em}Science \hspace{-.2em}and \hspace{-.2em}Engineering, \hspace{-.2em}Waseda \hspace{-.2em}University, \hspace{-.2em}Shinjuku, \hspace{-.2em}Tokyo, \hspace{-.2em}169-8555 \hspace{-.2em}Japan\\
$^2$Japan Atomic Energy Research Institute (JAERI), Tokai-mura, Naka-gun, Ibaraki, 319-1195 Japan}\\[.6em]
{\bf Abstract}
\end{center}
\baselineskip=.6cm
\begin{center}
\mbox{}\\[-2.2em]
\begin{minipage}{14.1cm}
We argue that the accuracy by which the isochron parameters of the decay $^{187}{\rm Re}\rightarrow \!^{187}\!{\rm Os}$ are determined by dating iron meteorites may not directly constrain the possible time-dependence of the decay rate and hence of the fine-structure constant $\alpha$.  From this point of view, some of the attempts to analyze the Oklo constraint and the results of the QSO absorption lines are re-examined.
\end{minipage}
\end{center}
\mbox{}\\[-1.8em]

Recently it has been claimed [\cite{olive}] that the long-lived beta decay
$^{187}{\rm Re}\rightarrow ^{187}\!\!{\rm Os}$ should constrain the
time-variability of the fine-structure constant $\alpha$ more strongly
than pointed out before [\cite{dyson}].  At the core of the argument,
however, the authors assumed that the limit of time variation of the
decay rate is of the same order as the accuracy by which the decay rate
itself is determined.  One of the purposes of this note is to re-examine this
hypothesis by checking the basic procedure of dating meteorites.  We
find that the decay rate measured at the {\em present time} is a {\em
time-average} of the {\em local} decay rate which can be time-dependent.
Obviously the rate of change, which is also time-dependent in
general, is entirely different from the accuracy of the average.  In
spite of this difference in principle, there might be a correlation in
practice, which is, however, to be tested observationally.

In order to show how we have reached this conclusion, we begin with a
brief account of how the decay rate of rhenium can be used to date 
iron meteorites for which recent analysis was carried out with an improved precision [\cite{science}].  The fundamental formulation will then be followed by the comments, (i)   the implication of direct measurement of today's value of the decay rate, (ii)  how Dyson's original idea is affected by our present knowledge of dating, and  (iii) a theoretical example in which the uncertainty of the averaged decay rate comes from the time dependence of the decay rate, illustrating that the relation between them can be complicated even if there is any in practice.

We also emphasize that the present issue attracts wide attention because
of the question if what is claimed to be the constraint from dating
meteorites can be understood consistently with other constraints on the
time-dependence of $\alpha$, from the Oklo phenomenon [\cite{oklo}] and
the QSO absorption lines [\cite{ww1},\cite{ww2}].   For this reason, the
second half of this note is devoted to a phenomenological re-examination
of the theoretical analyses in terms of the cosmological scalar field
[\cite{wett}--\cite{gardner}], in which fitting all of the observations on
the changing $\alpha$ has been attempted.  We find that the result is rather disappointing if we accept an important feature of the analysis of the QSO data; a rather ``flat" distribution of $\Delta\alpha/\alpha$ [\cite{ww1},\cite{ww2}].  We then suggest that an even better fit can be obtained if we do not impose the constraint from ${\rm Re}$-${\rm Os}$ decay.

Suppose a meteorite is formed at the time $t=t_1$, which is about 4.6 Gys ago, and will be called the ``meteorite time" in what follows.  After this time, the amounts of $^{187}{\rm Re}$ and the related $^{187}\!{\rm Os}$ are conveniently measured in terms of the ratios to the stable isotope $^{188}{\rm Os}$: $N_{\rm Re}\equiv ^{187}\!{\rm Re}/^{188}{\rm Os}$ and $N_{\rm Os}\equiv ^{187}\!\!{\rm Os}/^{188}{\rm Os}$, respectively.  The former would decay according to 
\beq
N_{\rm Re}(t) = A e^{-\lmd (t-t_1)},
\label{isoch-1}
\eeq
with the decay rate $\lmd$ assumed to be constant, for the moment. In practice, we consider a group of meteorites supposed to have started nearly at the same time, $t=t_1$.  The initial value $A=N_{\rm Re}(t_1)$ may be different from meteorite to meteorite due to chemical fractionations under different initial pressures and temperatures.  For each meteorite, the amount lost would appear as $^{187}{\rm Os}$;
\beq
N_{\rm Os}(t) = A \left( 1-e^{-\lmd (t-t_1)}\right) + B,
\label{isoch-2}
\eeq
where $B$ denotes the amount which had been present in the solar system before $t_1$ not necessarily due to the decay of $^{187}{\rm Re}$.  It seems reasonable to assume $B$ which is common throughout the group.  By eliminating $A$ from (\ref{isoch-1}) and (\ref{isoch-2}) we obtain
\beq
N_{\rm Os}(t) = S(t)N_{\rm Re}(t) + B,
\label{isoch-2-1}
\eeq
where the time-dependent $S(t)$ is given by
\beq
S(t) = e^{\lmd (t-t_1)}  -1,
\label{isoch-3}
\eeq
which starts with zero, increasing as $t-t_1$.

At the present time $t_0$ with $ t_0-t_1 \approx 4.6\times 10^9{\rm y}$, we measure $N_{\rm Re}(t_0)$ and $N_{\rm Os}(t_0)$, plotting the latter against the former which takes different values because of different $A$.  The result
should be a straight line (\ref{isoch-2-1}) with the common slope $S(t_0)$ given by (\ref{isoch-3}) with $t_0$ for $t$,  and the intercept $B$.  This plot is called the Re-Os ``isochron," as shown in Fig. 1 of [\cite{science}].  If the age $t_0-t_1$ is determined by other more accurate isochrons, U-Pb, for example, we may determine the rate $\lmd$.

Recently four groups (IIA, IIIA, IVA and IVB) of iron meteorites have been analyzed carefully [\cite{science}], determining the slopes with the accuracy of 0.5\%.  It was argued [\cite{olive}] that this accuracy is to be the same as the one by which $\Delta\lmd /\lmd$ of the time-variation of $\lmd$ is constrained.  They also exploited the theoretical relation [\cite{dyson}] 
\beq
\frac{\Delta\lmd}{\lmd}\approx -1.9\times 10^4\frac{\Delta\alpha}{\alpha},
\label{isoch-6}
\eeq
where the ``amplification factor" on the right-hand side comes from an exceptionally small $Q$-value, which makes the decay of $^{187}{\rm Re}$ particularly suitable to probe possible time-variability of $\alpha$ in the Coulomb-only approximation.  Imposing the same 0.5\% for the accuracy in determining the constant above on the left-hand side now interpreted as the one for the possible time-variation yields

\beq
\Biggl| \frac{\Delta\alpha}{\alpha} \Biggr|\:\lsim \:2.5\times 10^{-7},
\label{isoch-6-1}
\eeq
over the time span of $t_0-t_1$.  As it turns out this is roughly as strong as the constraint obtained from the Oklo phenomenon, and is nearly two orders of magnitude more stringent than the nonzero value indicated by the QSO result.  We will show, however, that the left-hand side of (\ref{isoch-6}) may not be 0.5\% mentioned above.

Suppose now $\lmd$ varies with time. Since the law of decay is described originally in terms of a differential equation, (\ref{isoch-1}) is modified to
\beq
N_{\rm Re}(t) = A \exp\left( \int_{t_1}^t \lmd(t')dt' \right).
\label{isoch-4}
\eeq
The expression (\ref{isoch-1}) with $t=t_0$ holds true if $\lmd$ is re-interpreted as an average:
\beq
\bar{\lmd}=\frac{1}{t_0-t_1}\int_{t_1}^{t_0} \lmd(t')dt'.
\label{isoch-5}
\eeq
All we can do from dating meteorites at the present time is to measure the average (\ref{isoch-5}), which masks any variation during the whole history between $t_1$ and $t_0$.  In this respect the method is different in principle from [\cite{oklo}] and [\cite{ww1},\cite{ww2}], in which the {\em past events} are directly probed.  We have no way to distinguish potentially  different functions $\lmd(t)$ as long as they give the same average value.   Any time-dependence in $\lmd(t)$ is allowed no matter how precisely $\bar{\lmd}$ is determined.  We are  going to illustrate the points by a few examples.

In order to detect time-dependence of $\lmd$, one of the most direct ways in principle is to find the present value $\lmd_0$, though it ought to be a difficult experiment because of an extremely small rate.  A recent accurate measurement [\cite{vitale}] provides
\beq
\lmd_0 = (1.68\pm 0.50)\times 10^{-11}{\rm y}^{-1},
\label{isoch-5-1}
\eeq
which is consistent with the value of $\bar{\lmd}= 1.67 \times
10^{-11}{\rm y}^{-1}$ used in [\cite{science}].  From the uncertainty of
3\% in (\ref{isoch-5-1}) follows that $\lmd(t)$ might have deviated
from $\bar{\lmd}$ {\em at least} by 3\%, or even by 6\%, sometime during the whole history of the meteorite, though depending on what time interval defines $\Delta\lmd/\lmd$.

We might assume the simplest linear approximation in which we have a straight line for $\lmd(t)$ which passes through $\bar{\lmd}$ at the middle, $(t_1 +t_0)/2$, giving $|\Delta\lmd/\lmd|\sim 3\times 10^{-2}$ at $t=t_1$.  Combining this with (\ref{isoch-6}) we find that $|\Delta\alpha /\alpha|\approx 1.5\times 10^{-6}$ toward 4.6 Gys ago.  This turns out to be much larger than (\ref{isoch-6-1}).  Notice that we could change the slope, the rate of time-change of $\lmd$ without changing the average, as long as we keep the middle point intact. A more precise determination of $\lmd_0$ will lower the lower-bound, but will never lower the upper-bound.

As a historical example of a non-linear function, let us re-evaluate Dyson's proposal [\cite{dyson}] who argued that if $\alpha$ toward the meteorite time was very large, and so was $\lmd$,  we would find no leftover $^{187}{\rm Re}$ around us today.  The contradiction can be avoided if a possible change of $\lmd(t)$, and of $\alpha(t)$, near the meteorite time is kept below certain limit.

Dyson's idea can be represented by a simplified step function,
\beq
\lmd(t)=\left\{
\begin{array}{ll}
\lmd_2, &\quad\mbox{for}\quad t_1 \leq t \leq t_2, \\
0, &\quad\mbox{for}\quad t_2 \leq t \leq t_0,
\end{array}
\right.
\label{isoch-7}
\eeq
where $t_2$ is an intermediate time between $t_1$ and $t_0$ while the
constant $\lmd_2$ is chosen to be higher than $\bar{\lmd}$, so that the
decay life-time defined by $\tau_2 = \lmd_2^{-1}$ is shorter than
$\bar{\tau}=\bar{\lmd}^{-1}$ (by 200 times in [\cite{dyson}]).  In order
to allow the largest possible $\lmd_2$ for a given interval $t_2-t_1$,
we chose in the second line the minimized $\lmd (t)$ satisfying the
natural condition $\lmd(t)\geq 0.$   We also add another narrow region
prior to $t_0$ for which $\lmd(t)=\bar{\lmd}$, assuming that today's
$\lmd$ equals $\bar{\lmd}$.  The time interval of this region can be
chosen very small, however, such that its effect to the integral will be negligibly small.

By requiring that the integral of (\ref{isoch-7}) gives $\bar{\lmd}$ correctly, we obtain the relation
\beq
(t_2-t_1)\lmd_2 = (t_0-t_1)\bar{\lmd}, \quad\mbox{or}\quad \frac{t_2-t_1}{\tau_2}= \frac{t_0-t_1}{\bar{\tau}}.
\label{isoch-9}
\eeq
By noticing that $\bar{\tau}= 6\times 10^{10}{\rm y}$ is much larger than $t_0-t_1$ in the last equation, we conclude
\beq
t_2-t_1 \ll \tau_2,
\label{isoch-11}
\eeq
implying that there was no time sufficiently long for the Re decay to be completed.  In other words, we still see rhenium around us no matter how short its decay life-time might be at the earliest time.  This shows that the established $\bar{\lmd}$ no longer allows a naive argument [\cite{dyson}].

As another exercise of the technique of a step function, let us start with a constant $\lmd(t)=\bar{\lmd}$ then introduce a local change $\lmd(t)= \lmd_2\equiv\bar{\lmd}+ \delta$ in the interval $t_1\leq t \leq t_2$.  We are replacing the second line, 0, of (\ref{isoch-7}) by $\bar{\lmd}$. The relative time-change $\Delta\lmd/\lmd$ may be defined by the change of $\lmd$ from ``today's" value $\bar{\lmd}$ divided by certain average of $\lmd(t)$.  The former is obviously $\delta$ while the latter is close to $\bar{\lmd}$ as long as $|\delta |\ll \bar{\lmd}$, hence
\beq
\frac{\Delta\lmd}{\lmd}= \frac{\delta}{\bar{\lmd}}.
\label{isoch-11-1}
\eeq
The uncertainty of the average $\bar{\lmd}$ will be determined by something different, like the isochron parameters.  It still seems interesting to suppose that possible inaccuracy in evaluating the contribution from $\delta$ causes $\Delta\bar{\lmd} \sim ((t_2-t_1)/(t_0-t_1))\delta$.  By combining this with (\ref{isoch-11-1}) we find
\beq
\frac{\Delta\bar{\lmd}}{\bar{\lmd}}\sim \frac{t_2-t_1}{t_0-t_1}\frac{\Delta\lmd}{\lmd}.
\label{isoch-12}
\eeq

This illustrates how the two kinds of relative changes can be different
from each other even if they share the same common origin, $\delta$ in
the range $t_1\leq t\leq t_2$.   In contrast to a {\em naive}
expectation for the basic equality between the two, as expressed in
[\cite{olive}], we realize that the relation depends on how $\lmd(t)$
behaves, through the ratio $(t_2-t_1)/(t_0-t_1)$ in our simplified
example.  We also point out that the numerator $t_2-t_1$ on the
right-hand side of (\ref{isoch-12}) is simply the interval of a nonzero change of $\lmd(t)$, independently of when it occurred, not necessarily at the earliest time.  A question might then be raised against a procedure in [\cite{wett}--\cite{gardner}], in which a constraint (\ref{isoch-6-1}) on $\Delta\alpha/\alpha$ is imposed right on the meteorite time.

We also note that, apart from the precise timing, the relation (\ref{isoch-12}), substituted from (\ref{isoch-6}) on the right-hand side, together with the isochron constraint $|\Delta\bar{\lmd}/\bar{\lmd}| \lsim 0.5\%$ may even derive  $|\Delta\alpha/\alpha| \sim 0.5\times 10^{-5}$, nearly the same as the weighted mean of the QSO result [\cite{ww1},\cite{ww2}], if we choose $(t_2-t_1)/(t_0-t_1)\sim 0.05$.

We have so far argued conceptually that the constraint
 (\ref{isoch-6-1}) may not be derived unambiguously, particularly at
 or near the meteorite time.  Now as a test of the  ``hypothesis'' in [\cite{olive}], we are going to show purely phenomenologically that this constraint is likely in conflict with the available QSO result.

For the later convenience, we define the fractional look-back time $u= 1-t/t_0$.  We then recall that the data points in [\cite{ww2}] (the ``fiducial sample," Fig. 8) distribute from $u=0.203 (z=0.229)$ to $u= 0.875 (z=3.666)$.  The meteorite time $u_{\rm met}=0.33$ is above the lowest data point, but certainly far below most of other data points.  All the 128 data points are summarized by
\beq
y \equiv \frac{\Delta\alpha}{\alpha}\times 10^{5} = -0.54\pm 0.12,
\label{isoch-13}
\eeq
with $\chi^2_{\rm rd}=1.06$.  Accommodating (\ref{isoch-6-1}), or $|y|= 0.025$, into the overall fit $y(u)$ might be achieved either by i) an exponential fall as $u$ decreases, or by ii) an oscillatory $y(u)$ with one of zeros nearly identified with $u_{\rm met}$.
The former choice has an advantage that other two important constraints, the Oklo constraint at $u_{\rm oklo}\approx 0.14 $ and $y(0)=0$ (by definition) would be complied almost automatically.

We thus start with assuming
\beq
y(u)= ae^{b(u-u_{\rm met})},
\label{isoch-14}
\eeq
where $b>0$ ($u$ is in the opposite direction with $t$) while $a=
y(u_{\rm met})$.  We expect that this is a behavior that 
simulates the dominant pattern of the fits attempted in [\cite{wett}--\cite{gardner}] reasonably well in the relevant time range.

We determine $a$ and $b$ by minimizing $\chi^2$ for the QSO data.  The
 reduced chi-squared $\chi^2_{\rm rd}= \chi^2/(128-2) = 1.07$ is obtained for $a=
-0.354$ and $b= 1.39$.  The above $\chi^2_{\rm rd}$ turns out to be
nearly the same as $1.06$ for the simple weighted mean (fit in terms of
a horizontal straight line, $y=-0.54$), while the size $a$ turns out
 more than an order of magnitude larger than the desired value $0.025$ for
(\ref{isoch-6-1}).  The corresponding fit is shown in Fig. 1, in which
 only the 13 binned data (see Fig. 8 of [\cite{ww2}]) is shown for a
 easier comparison with the fit.

\begin{minipage}[t]{7.4cm}
\vspace{-10em}
\baselineskip = 0.4cm
\epsfxsize=7.2cm
\epsffile{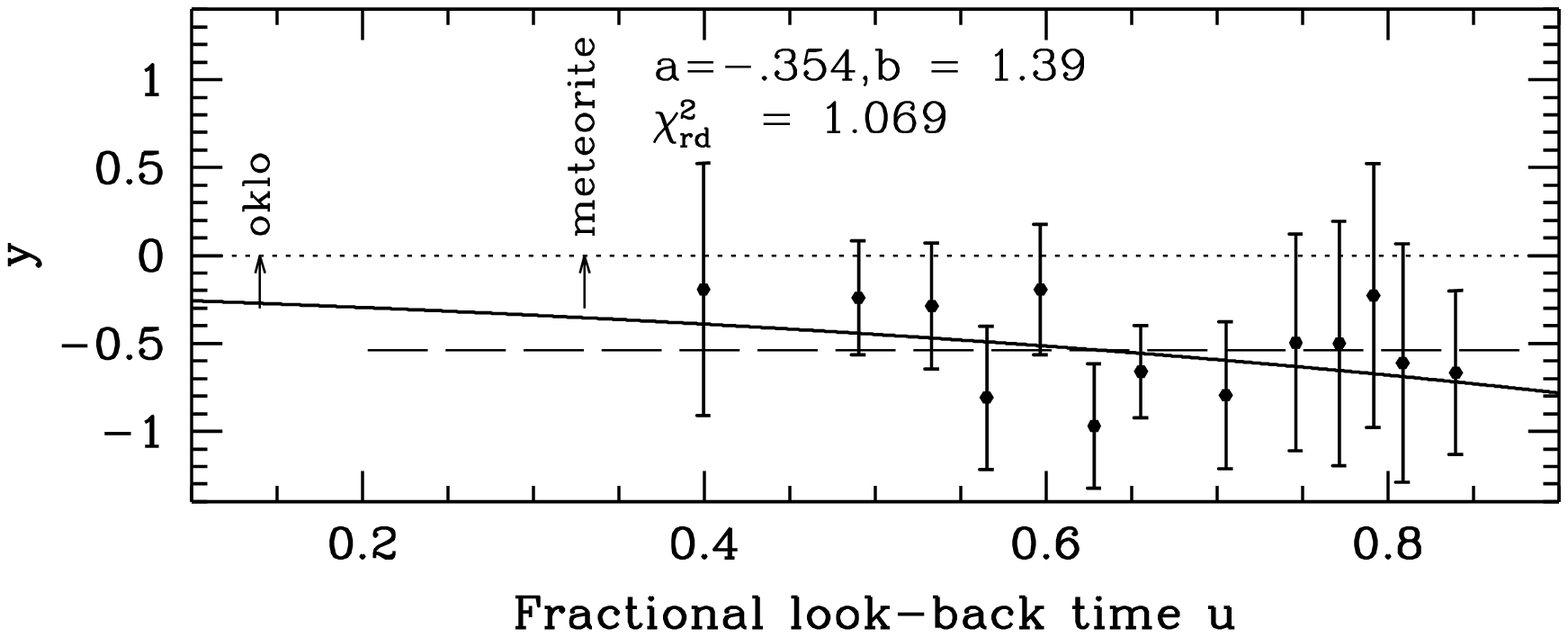}
\mbox{}\\[-2.1em]
\noindent
{\small Figure 1: An exponential fit (\ref{isoch-14}) to the QSO result
  [\cite{ww2}] shown as a binned fit, though  the minimizing $\chi^2$ itself has  been made with respect to the original 128 data points. The minimum
  $\chi^2_{\rm rd}=1.069$ is obtained for $a=-0.354$ and $b=1.39$.   The
  horizontal long-dashed line represents  the weighted mean $y=-0.54$.  A flat
  distribution of the data is obvious.  The Oklo time $u_{\rm oklo}=0.14$ 
  and the meteorite time $u_{\rm met}=0.33$  are shown 
  by upward arrows. 
}
\end{minipage}
\hspace{4mm}
\begin{minipage}[t]{7.4cm}
\vspace{-10em}
\baselineskip = 0.4cm
\epsfxsize=7.2cm
\epsffile{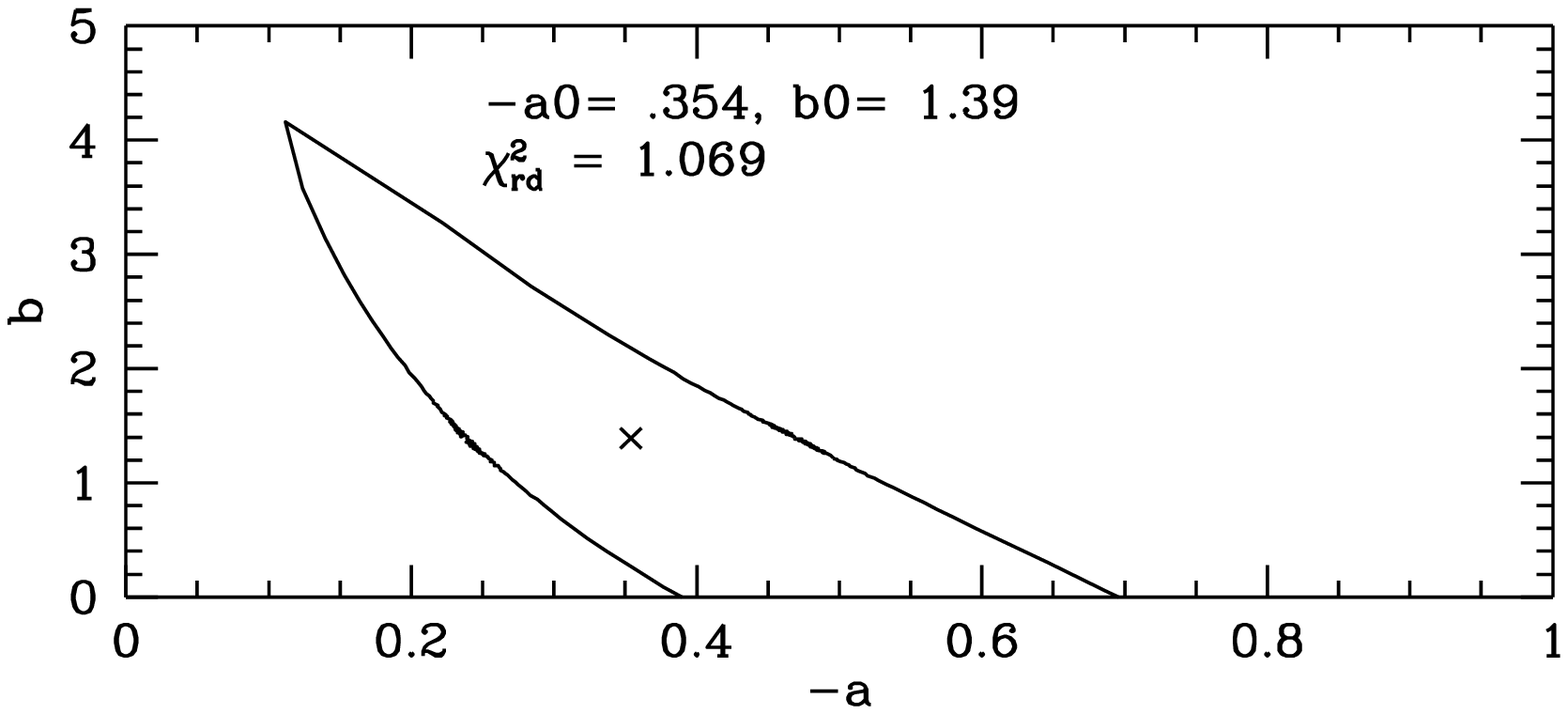}
\mbox{}\\[-2.1em]
\noindent
{\small Figure 2: The 68\% confidence region  around the minimized $\chi^2$ 
of the fit corresponding to  Fig. 1, as indicated by a cross.
}
\end{minipage}
\mbox{}\\

We also estimated the confidence region for 68\%, as shown in Fig. 2,
which shows that $a=-0.025$ is entirely out of the contour.  Around this
value of $a$, we find no 2-dimensional minimum of $\chi^2$.  We
tentatively fixed $a=-0.025$ and searched for $b$ to minimize $\chi^2$,
obtaining $b= 7.28$ with $\chi^2_{\rm rd}= 1.13$.  By comparing the 
corresponding fit shown in Fig. 3 with the previous Fig. 1, we find an obvious reason for a poorer agreement; the assumed small $-y(u_{\rm met})$ required a sufficiently large value of $b$, which in turn forced $-y$ to rise too much as $u$ increases, thus enhancing $\chi^2$.  In other words, a flat distribution of the QSO data points, as was seen also in the previous result [\cite{ww1}], dose not favor a
small $-y(u_{\rm met})$.  The same consequence
appears to be shared by the fits in [\cite{wett}] (with $E=12$), in
[\cite{anchor}] (Fig. 3) and in [\cite{gardner}] (Fig. 13).

\hspace*{.7cm}
\begin{minipage}{12.4cm}
\vspace{-9.9em}
\hspace*{2.2cm}
\baselineskip = 0.4cm
\epsfxsize=7.2cm
\epsffile{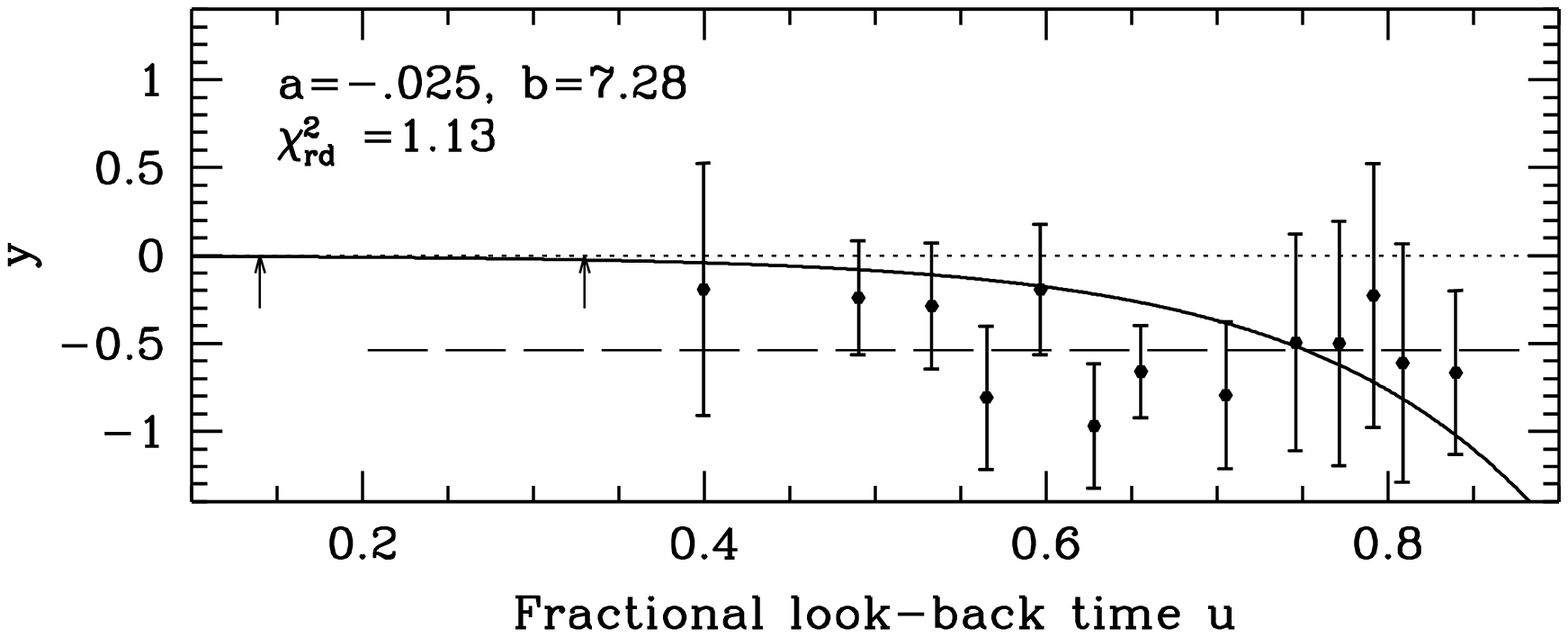}
\mbox{}\\[-1.5em]
\noindent
{\small Figure 3: A tentative fit with a fixed $a = -0.025$ but with $b=7.28$ that minimizes chi-squared, resulting in $\chi^2_{\rm rd}=1.13$.  Notice too sharp a fall of the curve toward the higher $u$.  
}
\end{minipage}
\\[1.5em]

We have further tried to see tentatively if the situation improves by replacing $u_{\rm met}\approx 0.33$ by $u_{\rm oklo}\approx 0.14$, but finding little changed basically.  We admit that the exponential suppression alone, like (\ref{isoch-14}), fails to accommodate the QSO result, as long as we accept the detailed behavior shown by [\cite{ww1},\cite{ww2}].

We have developed, on the other hand, a fit based on the damped-oscillator behavior as inspired by the scalar field in the scalar-tensor theory of gravity playing a crucial role in understanding the accelerating universe [\cite{cmbk}].  In [\cite{3prm}] we assumed a phenomenological function
\beq
y = a e^{bu}\sin \left( 2\pi \left( u-u_{\rm oklo} \right)/T \right),
\label{isoch-15}
\eeq
which is chosen to {\em vanish} at $u_{\rm oklo}$ as an approximate
expression for a small value $|y(u_{\rm oklo})|=10^{-2}-10^{-3}$.  The
smallest $\chi^2_{\rm rd}= 1.09$ was obtained for $a= 0.151, b= 2.4,
T=0.714$, fitting the QSO result nearly as good as the weighted-mean
with $\chi^2_{\rm rd}= 1.06$ [\cite{ww2}].  Notice that $b$ is smaller
than 7.28 as was used in Fig. 3.  This made it easier to be consistent
with  the flat
behavior of the QSO data toward the high-$u$ end.  Recall that we did
not appeal to a strong suppression of the exponential function for the
expected small value of $|y(u_{\rm oklo})|$.

We then find it  difficult, however, to satisfy the constraint
(\ref{isoch-6-1}) at $u_{\rm met}$ as well, because $u_{\rm met}-u_{\rm oklo}
\approx 0.19$ is much 
smaller than $T$, which is sufficiently large to keep the data distribution
``flat.''  Alternatively we may choose $u_{\rm met}$ as one of the zeros of $y$
and then trying to suppress  $y(u_{\rm oklo})$ sufficiently small.  This cannot be
immediately ruled out as long as we accept $b$ as large as 8.  This
choice, apart from the tendency of the cosmological solutions favoring
smaller $b$, makes it more difficult to comply another natural condition
$y(0)=0$.  If we follow the foregoing discussion {\em not} to appreciate the
 constraint (\ref{isoch-6-1}) at the meteorite time, we may propose a better, though still
phenomenological, fit to satisfy both of the constraints at $u=0$ and
$u_{\rm oklo}$ to be comfortably consistent with the QSO result, as will
be shown separately.

We finally point out that the constraints still from the Oklo and QSO
 results might be related to the evolution of the scalar field affecting 
directly the way of cosmological acceleration, through which we further
 look into the possibly changing $\alpha$ in the eras of CMB and the
 nucleosynthesis [\cite{cmbk}].

One of us (Y.F.) thanks Hisayoshi Yurimoto, Department of Earth and Planetary Sciences, Tokyo Institute of Technology, for his help to understand physics of meteorite dating.\\[1em]

\noindent
{\Large\bf References}
\begin{enumerate}
\item\label{olive}K.A. Olive, M. Pospelov, Y.-Z. Qian, A. Coc, M. Cass\'{e} and E. Vangioni-Flam, Phys. Rev. {\bf D66}, 045022 (2002).

\item\label{dyson}F.J. Dyson, Phys. Rev. Lett. {\bf 19}, 1291 (1967).

\item\label{science}M.I. Smoliar, R.J. Walker and J.W. Morgan, Science {\bf 271}, 1099 (1996).

\item\label{oklo}A.I. Shlyakhter, Nature {\bf 264}, 340 (1976); T. Damour and F.J. Dyson, Nucl. Phys. {\bf B480}, 37 (1996); Y. Fujii, A. Iwamoto, T. Fukahori, T. Ohnuki, M. Nakagawa, H. Hidaka, Y. Oura and P. M\"{o}ller, Nucl. Phys. {\bf B573}, 377 (2000); Proc. Int. Conf. on Nuclear Data for Science and Technology, 2001, hep-ph/0205206.

\item\label{ww1}J.K. Webb, M.T. Murphy, V.V. Flambaum, A. Dzuba, J.D. Barrow, C.W. Churchill, J.X. Prochaska and A.M. Wolfe,  Phys. Rev. Lett. {\bf 87}, 091301 (2001).

\item\label{ww2}M.T. Murphy, J.K. Webb, V.V. Flambaum, MNRAS, to be published, astro-ph/0306483.

\item\label{wett}C. Wetterich, Phys. Lett. {\bf B561}, 10 (2003).

\item\label{anchor}L. Anchordoqui and H. Goldberg, hep-ph/0306084.

\item\label{gardner}C.L. Gardner, astro-ph/0305080.

\item\label{vitale}M. Galeazzi, F. Fontanelli, F. Gatti and S. Vitale, Phys. Rev. {\bf C63}, 014302 (2000).

\item\label{cmbk}Y. Fujii and K. Maeda, {\sl The scalar-tensor theory of gravitation}, Cambridge University Press, 2003.

\item\label{3prm}Y. Fujii, to be published in Phys. Lett. B, astro-ph/0307263.

\end{enumerate}

\end{document}